\documentclass[seceq]{ptptex}
\usepackage{wrapft}
\usepackage{amsmath}
\usepackage{bm}

\title{
An Approach to $SU_q(2)$ Gauge Theory
}

\author{
Shigefumi \textsc{Naka}$^{1,}$\footnote{E-mail:naka@phys.cst.nihon-u.ac.jp}, Akiyuki \textsc{Kinouchi}$^{1,}$\footnote{E-mail:akitama@phys.cst.nihon-u.ac.jp} and Haruki \textsc{Toyoda}$^{2}$\footnote{E-mail:toyoda@gaea.jcn.nihon-u.ac.jp} 
}
\inst{
$^1$Department of Physics, College of Science and Technology
Nihon University,\\ 1-8-14 Kanda-Surugadai Chiyoda-ku Tokyo, Japan
\\
$^2$Junior College, Funabashi Campus, Nihon University,\\ 7-24-1 Narashinodai Funabashi-shi Chiba, Japan
}
\abst{
In the usual approach to q-deformed gauge theories, the gauge fields are required to be non-local or non-commutative one's. If we introduce, however, an extended product, which we call \lq\lq $\star$-product\rq\rq, among the generators of a q-deformed Lie group, the deformed group can be reduced to a ordinary Lie group under the $\star$-product. According to this line of approach, we try to construct a $[SU_q(2)\times U(1)]_\star$, a $SU(2)\times U(1)$ analogue under the $\star$-product, gauge theory. In this gauge theory with the $\star$-product, the $U(1)$ symmetry is naturally incorporated into the $SU(2)$ symmetry. We also study the symmetry breaking by the Higgs mechanism associated with $J=\frac{1}{2}$ and $J=1$ representations of $SU_q(2)$ algebra, and show that the mixing angle between the $SU(2)$ and $U(1)$ gauge fields is determined uniquely in a tree level.
}

\begin{document}
\maketitle

\section{Introduction}

The \lq\lq q-deformations\rq\rq are mappings of dynamical systems through modifications of commutation relations associated with underlying quantum groups or non-commutative geometries\cite{q-deform},\cite{Wess},\cite{Manin}\cite{q-spacetime}. It is known that the deformations cause quite change of dynamical systems or symmetry groups. Because of this reason, many studies have been made for the q-deformation of various dynamical systems and symmetry groups\cite{Quantum-Groups}. Its applications for field theories were also made extensively from several points of view: conformal field theories\cite{Faddeev}, fields with deformed internal symmetry\cite{Mesref}, non-local fields with deformed extra-coordinates\cite{Naka-Toyoda}, and so forth.

In particular, the deformation of gauge groups is expected to give a new insight into the symmetry breaking, since the symmetry under a deformed Lie group sometimes breaks the one under the usual Lie group before deformation\cite{q-gauge}. When we apply this idea to gauge theories, there appear to be two approaches handling gauge fields. One is to start from a matrix representation of a deformed gauge Lie group with non-commutative matrix elements\cite{Finkelstein}. In this case, gauge field components associated with a deformed gauge Lie group become non-commutative one's. Another approach is to start with a matrix representation of a deformed Lie algebra. For example, the $SU_q(2)$ generators $\{J_i\},(i=\pm,3)$ are required to satisfy $[J_{+},J_{-}]=[2J_3]\neq 2J_3$. Here $[2J_3]$ is a function of $J_3$ including one parameter $q$; and so, according to this line of approach, we need many components of gauge field\cite{Arefeva-Volovich} corresponding to $(J_3)^n$, though the number of generators is three.
 
The purpose of this paper is to study the other line of approach to a q-deformed gauge theory. Considering the application to the electroweak gauge theory, we focus our attention on a $SU_q(2)$ gauge theory. Then we can show that the $SU_q(2)$ generators satisfies an algebra as if the gauge group is the $SU(2)\times U(1)$ under a modified product between generators, which we call \lq\lq $\star$-product\rq\rq. Then, since the $U(1)$ symmetry is included in the gauge group in a non-trivial manner, the Weinberg angle is determined in tree level, although the numerical value is not close to phenomenological one.

In the next section, we first summarize the representation of standard $SU_q(2)$ algebra. Then we discuss the $\star$-product which modifies the $SU_q(2)$ algebra as if it is a rank two algebra. The section 3, is devoted to the construction of modified electroweak gauge theory associated with the $[SU_q(2)\times U(1)]_\star$ symmetry in our notation. There, the Higgs fields are treated as components of $J$-dimensional representation of $SU_q(2)$, to which the substitution $J\rightarrow 1/2$ is taken after all calculation; then, we can pull out symmetry breaking effects in a unique way. In addition to this, we also attempt to construct a model associated with triplet Higgs fields, which belong to $J=1$ representation of $SU_q(2)$. In this case, the mixing between $SU(2)$ and $U(1)$ is a direct result of a non-trivial deformation $q\neq 1$. \S 4 is the summary and discussion. We also give a short review for the representation of $SU_q(2)$ algebra in Appendix A.

\section{$SU_q(2)$ with a $\star$-product}

The algebra of $SU_q(2)$ is defined by
\begin{align}
 [J_3,J_\pm] &= \pm J_\pm , \label{suq2-1}\\
 [ J_{+},J_{-} ] &= [ 2J_3 ] \label{suq2-2} ,
\end{align}
where $[x]$ is a function of $x$ specified by one parameter $q$:
\begin{equation}
[x]=\frac{q^x-q^{-x}}{q-q^{-1}},
\end{equation}
which tends to $x$ according as $q\rightarrow 1$. The second Casimir operator of this algebra is given by
\begin{equation}
 \bm{J}^2=J_{-}J_{+}+[J_3][J_3+1]=J_{+}J_{-}+[J_3][J_3-1]. \label{Casimir}
\end{equation}
Then, the basis of $J$-dimensional representation of this algebra have a q-deformed structure of $SU(2)$ one such that 
\begin{align}
 \bm{J}^2|J,M\rangle &= [J][J+1]|J,M\rangle,~(J=0,1/2,1,\cdots), \label{eigen-1}\\
 J_3|J,M\rangle &= M|J,M\rangle, ~(M=-J,-J+1,\cdots, J). \label{eigen-2}
\end{align}
The orthnormalized states $|J,M\rangle,~(M=-J,-J+1,\cdots,J)$ can be constructed from the lowest $J_3$ state $|J,-J\rangle$ with the normalization $\langle J,-J|J,-J \rangle=1$ by 
\begin{equation}
 |J,M\rangle = \frac{(J_+)^{J+M}}{\sqrt{([J][J+1]-[M][M-1])!}}|J,-J\rangle , \label{normalization}
\end{equation}
where $f(M)! \equiv f(-J)f(-J+1)\cdots f(M)$. Then, it is not difficult to verify that
\begin{align}
 J_+|J,M\rangle =\sqrt{[J][J+1]-[M][M+1]}|J,M+1\rangle , \\
 J_-|J,M\rangle =\sqrt{[J][J+1]-[M][M-1]}|J,M-1\rangle .
\end{align}
In particular, $J_i$ becomes respectively $\frac{1}{2}\sigma_i,(i=\pm,3)$ for $J=1/2$. Here $\sigma_i$'s are Pauli matrices.

In spite of the similarity between $SU(2)$ and $SU_q(2)$, it is not easy to set up a $SU_q(2)$ gauge field theory, since the right-hand side of eq.(\ref{suq2-2}) contains infinite higher powers of $J_3$. Then, the form $W_\mu^iJ_i\equiv \frac{1}{\sqrt{2}}(W^+J_+W^-J_-)+W^3J_3$ is not closed under the unitary transformation by $U(\delta\theta)=\exp\{i\delta\theta^iJ_i\}\simeq 1+i\delta\theta^iJ_i$; indeed we can verify that
\begin{equation}      
 U^\dag W_\mu^iJ_i U = W_\mu^iJ_i+\frac{i}{\sqrt{2}}\left(\delta\theta^{[+}W_\mu^{3]}J_{+} - \delta\theta^{[-}W_\mu^{3]}J_{-} -\frac{1}{\sqrt{2}}\delta\theta^{[+}W_\mu^{-]}[2J_3] \right)~,
\end{equation}
and the right-hand of this equation can not be written in the form $(W_\mu^i+\delta W_\mu^i)J_i$.

One way to get rid of this difficulty is to introduce infinite number of gauge fields such as $\frac{1}{\sqrt{2}}(W^{+}J_{+}+W^{-}J_{-})+\sum_n W^{(n)}(J_3)^n$. In this case, however, we face another problem to explain infinite unknown components of gauge fields. In what follows, we try another approach to a gauge theory based on $SU_q(2)$ symmetry. The key is that there is a function $\eta(J_3)$ satisfying
\begin{equation}
 \eta J_{+}J_{-}-J_{-}J_{+}=\alpha J_0+\beta J_3, \label{eta-commutator}
\end{equation}
where $J_0$ is the unit operator and $\alpha, \beta$ are operator depending only on the second Casimir invariant; that is, that these may be functions of $J$ in $2J+1$ dimensional representation of $SU_q(2)$ algebra.

Since eq.(\ref{Casimir}) yields $J_\pm J_\mp=[J][J+1]-[J_3][J_3 \mp 1]$, the $\eta$ can be formally solved as
\begin{equation}
 \eta(J_3)=\frac{\alpha J_0 + \beta J_3 +([J][J+1]-[J_3][J_3+1])}{[J][J+1]-[J_3][J_3-1]}~. \label{eta-1}
\end{equation}
The denominator of this expression contains $0$ at $J_3=-J$; and so, we require that $J_3=-J$ is the same order of zero in the numerator too. In addition to this, we require $\eta \rightarrow 1$ according to $q \rightarrow 1$. These two requirements determine $\alpha$ and $\beta$ such that
\begin{equation}
 \alpha=2J-[J][J+1]+[J][J-1]=2J-[2J]~~~\mbox{and}~~~\beta=2~.
\end{equation}
Substituting these expressions for eq.(\ref{eta-1}), the $\eta$ is determined as
\begin{equation}
 \eta(J_3)=\frac{2(J+J_3)+[J][J-1]-[J_3][J_3+1]}{[J][J+1]-[J_3][J_3-1]}~. \label{eta-2}
\end{equation}
Therefore, the commutator (\ref{eta-commutator}) associated with $\eta$ will reduce to the ordinary $SU(2)$ commutator $[J_+,J_-]=2J_3$ in the limit $q\rightarrow 1$. We note that in the exceptional case $J=\frac{1}{2}$, this reduction is realized even for $q\neq 1$.

Now, eq.(\ref{eta-commutator}) suggests to introduce a new product between $SU_q(2)$ generators such as
\begin{equation}
 J_i\star J_j \equiv R_{ij}^{kl}J_k J_l ,~~(~R_{ij}^{kl}=\delta_i^k\delta_j^l+(\eta-1)\delta_i^{+}\delta_j^{-} ~)
\end{equation}
; that is, $J_{+}\star J_{-}=\eta J_{+}J_{-}$ for $i=+,j=-$ and $J_i\star J_j=J_i J_j$ otherwise. For this $\star$-product, one can verify easily the associative law $J_i\star (J_j\star J_k)=(J_i\star J_j)\star J_k$ in addition to the distributive law, which is obvious by definition. Then, with this $\star$-product, the eqs.(\ref{suq2-1}) and (\ref{eta-commutator}) can be written as
\footnote{
The operator $\eta$ is determined for each irreducible representatio of $SU_q(2)$ depending on $J$. However, since ${\bm J}^2=[J][J+1]=[J+\frac{1}{2}]^2-[\frac{1}{2}]^2$, we may read $J$ as the operator $J=(\log q)^{-1}\sinh^{-1}\{\sinh(\log q)\sqrt{{\bm J}^2+[\frac{1}{2}]^2} \}-\frac{1}{2}$. In this sence, these equations can be understood as operator equations.
}
\begin{align}
 [J_3,J_{\pm}]_\star &=\pm J_{\pm}~, \label{*algebra-1}\\
 [J_{+},J_{-}]_\star &=\alpha J_0+2 J_3~, \label{*algebra-2}
\end{align}
where $[A,B]_\star \equiv A\star B-B\star A$. Equations (\ref{*algebra-1}) and (\ref{*algebra-2}) imply that $J_{\pm}$ and $J_3$ form a $U(2)$ like algebra under the $\star$ commutator, in which the $U(1)$ generator $J_0$ is included in a non-trivial manner. It should be noticed that the $J_0$ in the right-hand side of eq.(\ref{*algebra-2}) is necessary because of $Tr(J_{+}\star J_{-})\neq Tr(J_{-}\star J_{+})$.  

The commutator with the $\star$-product causes the transformations among $J_i$ and $J_0$ in the following sense: the unitary operator $U(\delta\theta)=\exp\{i\delta\theta^A J_A\}\simeq 1+i\delta\theta^A J_A,~(A=i,0)$ with infinitesimal parameters $\delta\theta^A$ allow us to calculate $U^\dag\star (W^A J_A)\star U=(W+\delta W)^A J_A$. This means that we have to define the transformation of states in their products by
\begin{equation}
 \delta_\star|\Phi\rangle=\star i\delta\theta^A J_A|\Phi\rangle~~\mbox{and}~~\delta_\star\langle \Phi|=-\langle\Phi|i\delta\theta^A J_A\star
\end{equation}
Then it holds obviously that $\delta_\star\langle\Psi|\Phi\rangle=0$ and $\delta_\star\langle\Psi|J_B|\Phi\rangle=i\delta\theta^A\langle\Psi|[J_A,J_B]_\star|\Phi\rangle$.

Now, for the latter purpose, we here rewrite eqs.(\ref{*algebra-1}) and (\ref{*algebra-2}) as
\begin{align}
 [{\cal J}_3,{\cal J}_\pm ]_\star &= \pm {\cal J}_\pm ~, \label{cal-1}\\
 [{\cal J}_{+},{\cal J}_{-}]_\star &= 2{\cal J}_3 ~, \label{cal-2}
\end{align}
where
\begin{equation}
 {\cal J}_\pm =J_\pm ~~~\mbox{and}~~~{\cal J}_3=J_3+\frac{\alpha}{2}J_0~. \label{redefinition}
\end{equation}
The $\star$-algebra (\ref{cal-1}) and (\ref{cal-2}) are, then, nothing but those of $SU(2)$ by reading $[~,~]_\star \rightarrow [~,~]$. In the generators $\{ {\cal J}_a,J_0 \}$, however, the ${\cal J}_3$ and $J_0$ are not linearly independent because of $Tr({\cal J}_3J_0)\neq 0$. To get a linearly independent set of generators, let us introduce a new crew 
\begin{equation}
 {\cal J}_0=J_0-\frac{N_0}{N_3}\frac{\alpha}{2}J_3~,
\end{equation} 
where
\begin{align}
 N_0 &=2Tr(J_0^2)=2(2J+1)~, \\
 N_3 &=2Tr(J_3^2)=\frac{2}{3}(2J+1)J(J+1)~.
\end{align}
Then  $Tr({\cal J}_3{\cal J}_0)=0$ holds obviously, and ${\cal J}_0$ adds remaining algebra to (\ref{cal-1}) and (\ref{cal-2}):
\begin{align}
 [{\cal J}_0,{\cal J}_\pm ]_\star &= \mp k_J {\cal J}_\pm ~, ~~\left( k_J=\frac{N_0}{N_3}\frac{\alpha}{2} \right)  \label{cal-3}\\
 [{\cal J}_0,{\cal J}_3]_\star &= 0 ~. \label{cal-4}
\end{align}
If it is necessary, we may normalize these generators so that $Tr(\tilde{\cal J}_{+}\tilde{\cal J}_{-})=1$ and $Tr(\tilde{\cal J}_3^2)=Tr(\tilde{\cal J}_0^2)=\frac{1}{2}$ hold. This can be done by putting
\begin{equation}
 \tilde{\cal J}_\pm=\frac{1}{\sqrt{N_{\pm}}}{\cal J}_\pm,~\tilde{\cal J}_3=\frac{1}{\sqrt{{N}^\prime_3}}{\cal J}_3,~\tilde{\cal J}_0=\frac{1}{\sqrt{N^\prime_0}}{\cal J}_0,
\end{equation}
where
\begin{align}
 N_{\pm} &= Tr({\cal J}_{+}{\cal J}_{-})=[J][J+1](2J+1)-\frac{q+q^{-1}}{(q-q^{-1})^2}( [2J+1]-(2J+1) ) \\
 & {}\hspace{23mm}=\frac{(2J)[2J+2]-[2J](2J+2)}{(q-q^{-1})^2}, \\
  N^\prime_3 &= 2Tr({\cal J}_3^2)=N_3\left\{1+\frac{N_0}{N_3}\left(\frac{\alpha}{2}\right)^2\right\}, \\
 N^\prime_0 &= 2Tr({\cal J}_0^2)=N_0\left\{1+\frac{N_0}{N_3}\left(\frac{\alpha}{2}\right)^2\right\},
\end{align} 
from which we have $\frac{N_0^\prime}{N_3^\prime}=\frac{N_0}{N_3}$. We also note that these normalization factors $N_0^\prime,N_3^\prime$, and $N_\pm$ tend respectively to $N_0,N_3$, and $\frac{1}{3}(2J+1)J(J+1)$ in the limit $q\rightarrow 1$.

Equations (\ref{cal-1}),(\ref{cal-2}),(\ref{cal-3}), and (\ref{cal-4}) says that the  ${\cal J}_A,~(A=a,0)$ form a closed algebra under the $\star$ commutator. In what follows, we shall call the symmetry associated with this algebra as $[SU_q(2)\times U(1)]_\star$ symmetry. 

\section{$[SU_q(2)\times U(1)]_\star$ gauge symmetry}

We are ready for formulating a gauge theory based on the $[SU_q(2)\times U(1)]_\star$ symmetry. The gauge fields in this case can be introduced associated with the covariant derivative defined by 
\begin{align}
 D_\mu(W) &=\partial_\mu +ig\left\{ \frac{1}{\sqrt{2}}(W^+_\mu\tilde{\cal J}_+ +W_\mu^-\tilde{\cal J}_-)+W^3_\mu\tilde{\cal J}_3 +W^0_\mu\tilde{\cal J}_0 \right\} \label{covariant-1} \\
 &=\partial_\mu +ig \frac{1}{\sqrt{2}} \left( \bar{W}^+_\mu{\cal J}_+ +\bar{W}_\mu^-{\cal J}_- \right)  + ig_3\bar{W}^3_\mu{\cal J}_3+ig_0W^0_\mu S , \label{covariant-2}
\end{align}
where $S=\frac{Y}{2}J_0$, and $Y$ is a J-dependent parameter representing a hypercharge of the matter field, to which $D_\mu$ operates. Further, we have put
\begin{align}
 g_0 &=g\frac{2}{Y}\frac{1}{\sqrt{N_0^\prime}}\left\{1+\frac{N_0}{N_3}\left(\frac{\alpha}{2}\right)^2\right\}=\frac{2g}{Y\sqrt{N_0}}\sqrt{1+\frac{N_0}{N_3}\left(\frac{\alpha}{2}\right)^2}, \\
 g_3 &= g\sqrt{\frac{1}{N_3^\prime}+\frac{1}{N_0^\prime}\left(\frac{N_0}{N_3}\frac{\alpha}{2}\right)^2}=\frac{g}{\sqrt{N_3}}~, \label{couplings}
\end{align}
and
\begin{equation}
 \left( \bar{W}_\mu^\pm,\bar{W}_\mu^3 \right) = \left\{ \frac{1}{\sqrt{N_\pm}}W_\mu^\pm, \frac{g}{g_3}\left(\frac{1}{\sqrt{N_3^\prime}}W_\mu^3-\frac{1}{\sqrt{N_0^\prime}}\frac{N_0}{N_3}\frac{\alpha}{2}W_\mu^0 \right) \right\}. \label{W-bar}
\end{equation}
Here, the $\bar{W}^3$ is normalized so that the transformation from $W^3$ to $\bar{W}^3$ becomes a rotation in $(W^3,W^0)$ space. 

The covariant derivative (\ref{covariant-2}) implies that the $(\bar{W}^\pm,\bar{W}^3)$ are crew of $SU(2)$ gauge fields in the $[SU_q(2)\times U(1)]_\star$ symmetry, though the $SU(2)$ gauge symmetry is already broken due to $g\neq g_3$. This means that the recombinant gauge fields $(\bar{W}_\mu^\pm,\bar{W}_\mu^{(3)}=\frac{g_3}{g}\bar{W}^3_\mu)$ and $W^0$ transform as ordinary $SU(2)\times U(1)$ gauge fields under the unitary transformation
\begin{equation}
 U \simeq 1+i\left\{ \frac{1}{\sqrt{2}}\left(\delta\theta^+{\cal J}_+ +\delta\theta^-{\cal J}_-\right)+\delta\theta^3{\cal J}_3 +\delta\theta^0 S \right\} 
\label{unitary}
\end{equation}
with the $\star$-product. Namely, we can obtain $U^\dag \star D_\mu(W)\star U=D_\mu(W+\delta W)$, where
\begin{equation}
  \delta \bar{W}^\pm_\mu =\frac{1}{g}\partial_\mu\delta{\theta}^\pm \pm i\delta{\theta}^{[\pm} \bar{W}_\mu^{(3)]}~, ~~
  \delta\bar{W}^{(3)}_\mu =\frac{1}{g}\partial_\mu\delta{\theta}^3 - i\delta{\theta}^{[+}\bar{W}_\mu^{-]}~, \label{SU(2)}
\end{equation}
and
\begin{equation}
  \delta{W}^0_\mu =\frac{1}{g_0}\partial_\mu\delta{\theta}^0. \label{U(1)}
\end{equation}
Then, the field strengths for this $SU(2)\times U(1)$ symmetry can be defined by
\begin{equation}
 F_{\mu\nu}=\frac{1}{ig}[D_\mu,D_\nu]_\star =\frac{1}{\sqrt{2}}\left(\bar{F}_{\mu\nu}^{+}{\cal J}_{+} + \bar{F}_{\mu\nu}^{-}{\cal J}_{-}\right)+\bar{F}_{\mu\nu}^{3}{\cal J}_3-\frac{g_0}{g}F_{\mu\nu}^0 S,
\end{equation}
where
\begin{align}
\bar{F}_{\mu\nu}^\pm &=\partial_{[\mu}\bar{W}_{\nu]}^\pm \pm ig \bar{W}_{[\mu}^{(3)}\bar{W}_{\nu]}^\pm ~, \\
\bar{F}_{\mu\nu}^{3} &=\partial_{[\mu}\bar{W}_{\nu]}^{(3)}+ig\bar{W}_{[\mu}^{+}\bar{W}_{\nu]}^{-} ~, \\
F_{\mu\nu}^0 &=\partial_{[\mu}W_{\nu]}^0 ~.
\end{align}
Therefore, we can write down the action
\begin{equation}
 {\cal L}_W=-\frac{1}{2}\sum_{a,b=\pm,3}g_{ab}\bar{F}^a_{\mu\nu}\bar{F}^{b\mu\nu}-\frac{1}{4}F_{\mu\nu}^0F^{0\mu\nu}~,\left(g_{ab}=Tr(\tilde{\cal J}_a\tilde{\cal J}_b)\right),  \label{action}
\end{equation}
which is invariant under the transformations (\ref{SU(2)}) and (\ref{U(1)}), although the $SU(2)$ invariance is not realized for $(\bar{W}^\pm,\bar{W}^3)$ but for $(\bar{W}^\pm,\bar{W}^{(3)})$.

Next, let us consider a gauge-Higgs system to evaluate the effect of spontaneous symmetry breaking in this q-deformed gauge theory. The results depend on the dimension of $SU_q(2)$ representations. In the following, we shall discuss typical two cases. \vspace{4mm}

\noindent
{\bf case i)} \vspace{2mm}\\
The Higgs fields in the standard electroweak theory belongs to a $J=\frac{1}{2}$ isospin doublet $\phi=\binom{\phi^+}{\phi^0}$, to which the charge operator is defined by $Q=J_3+S$ with $Y_{H}=1$. The leptonic fields are, then, consisting of the $J=\frac{1}{2}$ left-handed fermions $\psi_L=\binom{\nu_e}{e}$ and the $J=0$ right-handed electron $\psi_R=(e)_R$. The charge operators of $\psi_L$ and $\psi_R$ fields again satisfy $Q=J_3+S$ by assigning $Y_L=-1$ and $Y_R=-2$ respectively. Then the Yukawa interaction $\bar{\psi}_R\phi^\dag\psi_L$ is invariant under the $SU(2)$ by $\{{\cal J}_i\},(i=\pm,3)$ generators and the $U(1)$ by $S$ generator.

As discussed in \S2, however, the $J=\frac{1}{2}$ representation is an exceptional case for Lie $SU_q(2)$ algebra; then, the generators $J_i$ of $SU_q(2)$ are reduced to those of $SU(2)$. To make clear the $q$-dependence in the spontaneous symmetry breaking, it is worthwhile to discuss the Higgs fields belonging to the $J$-dimensional representation of $SU_q(2)$ on a temporary basis. 

As usual, the action for the Higgs field can be written as 
\begin{equation}
 {\cal L}_H=\frac{1}{2}\langle D_\mu\phi|D^\mu\phi\rangle -\frac{\lambda}{2}\left(\langle\phi|\phi\rangle -\frac{v^2}{2}\right)^2 \label{Higgs}
\end{equation}
with the covariant derivative operators (\ref{covariant-1}) or (\ref{covariant-2}) belonging to $J$-dimensional representation of $SU_q(2)$. Further, the bracket $\langle \cdot\cdot|\cdot\cdot \rangle$ is the inner product between two states in the $J$-dimensional representation space of $SU_q(2)$. The action is invariant under the unitary transformation (\ref{unitary}) with the $\star$-product associated with the gauge field transformations $W \rightarrow W-\delta W$.

Under these preparations, we can evaluate the $q$ dependencies of the Weinberg angle $\theta$ and the mass ratio $M_W/M_Z$ in this framework. Substituting, first, the rotation
\begin{align}
 \bar{W}_\mu^{3} &=A_\mu\sin\theta + Z_\mu\cos\theta \label{rotation-1} \\
 W_\mu^0 &= A_\mu\cos\theta-Z_\mu\sin\theta \label{rotation-2}
 \end{align}
for $\bar{W}^3,W^0$ terms in Eq.(\ref{covariant-2}), we can find
\begin{equation}
 g_3 \bar{W}_\mu^{3}{\cal J}_3+g_0W_\mu^0 S=eQA_\mu+\frac{e}{\sin\theta\cos\theta}\left\{ {\cal J}_3-Q\sin^2\theta \right\}Z_\mu, \label{A-Z}
\end{equation}
where $e=g_3\sin\theta$ and $g_0=\frac{e}{\cos\theta}(1-\frac{\alpha}{Y_H})$. Hence, taking $\left(\frac{g_3}{g_0}\right)^2 =\left(\frac{Y_H}{2}\right)^2\frac{N_0}{N_3}\frac{1}{1+\frac{N_0}{N_3}\left(\frac{\alpha}{2}\right)^2}$ into account, one can obtain
\begin{equation}
 \sin^2\theta =\frac{1}{1+\left(\frac{g_3}{g_0}\right)^2\left(1-\frac{\alpha}{Y_H}\right)^2} = \frac{1}{1+\left(\frac{N_0}{N_3}\right)\frac{1}{1+\frac{N_0}{N_3}\left(\frac{\alpha}{2}\right)^2}\left(\frac{Y_H}{2}-\frac{\alpha}{2}\right)^2}. \\ \label{sin-theta}
\end{equation}
It should be noted that the $\tan\theta=\frac{g_0}{g_3}(1-\frac{\alpha}{Y_H})^{-1}$ does not coincide with the standard form $\frac{g_0}{g_3}$ owing to the effect of $\alpha\neq 0$. 

Next, to evaluate the mass ratio $M_W/M_Z$ by the Higgs mechanism, let us assume that the potential for $\phi$ fields pick out the vacuum expectation value $\phi_0$ satisfying $Q\phi_0=0,(i.e.,~J_3\phi_0=-S\phi_0~{\rm or}~{\cal J}_3\phi_0=\frac{\alpha-Y_H}{2}\phi_0)$ and $\langle \phi_0|\phi_0\rangle=\frac{v^2}{2}$, then the photon field $A_\mu$ becomes obviously massless. Simultaneously with this, by taking $\{J_\pm, J_\mp\} \phi_0=(2[J][J+1]+[2S]-2[S][S+1])\phi_0$ into account, the $Z$ and $W$ couplings in the kinetic term of Higgs field give rise to the mass terms for $Z$ and $W$ such as
\begin{align}
 M_Z &= \frac{ev}{\sin\theta\cos\theta}\left| \frac{\alpha-Y_H}{2} \right|~, \label{Z-boson} \\
 M_{\bar{W}} &= \frac{ev}{2\sin\theta}\frac{g}{g_3}\sqrt{2[J][J+1]+[2S]-2[S][S+1]}~. \label{W-boson}
\end{align}

The equations (\ref{sin-theta})$\sim$(\ref{W-boson}) yield the value $\sin^2\theta$ and the ratio $M_{\bar{W}}/M_Z$ in the limit $J \rightarrow 1/2$. In this limit, one can verify even for $q\neq 1$ that $\alpha=0,~N_3^{\prime}=N_3=1,~N_0^{\prime}=N_0=4$, and $N_\pm=1$ respectively. Therefore, putting $Y_{H}=1$, we finally obtain
\begin{equation}
 \sin^2\theta = \frac{1}{2} = 0.5~~~~\mbox{and}~~~~M_W=M_Z\cos\theta~.
\end{equation}
These results are unrealistic from a view point of phenomenology; and, the results are expected from eq.(\ref{covariant-2}) in advance, since $g_3$ and $g_0$ terms are reduced to $g(\bar{W}_\mu^3{\cal J}_3+ W_\mu^0S)$ in the limit $J\rightarrow 1/2$ with $Y_{H}=1$. \vspace{4mm}

\noindent
{\bf case ii)} \vspace{2mm}\\
The next is a toy model, which assigns a triplet Higgs fields $\Phi=(\phi^{+},\phi^{0},\phi^{-})^T$ to the $J=1$ representation of $SU_q(2)$ so that $Q=J_3$ in this case. We also assign leptonic fields within two generations to a triplet\cite{triplet} $\psi=(\mu^{+},\nu,e^{-})^T$ with $\mu_L=\nu_e$ and $\nu_R=\nu_\mu^{c}$. Even in this case, we can define a $U(1)$ hypercharge $Y_H(\neq 0)$ for $\Phi$ field, which is a parameter independent of $Q$. In other words, the usual relation $Q=J_3+S$ is not applied to those matter fields. 

Now, the covariant derivative of the Higgs fields with $(\bar{W}^{\pm}_\mu,\bar{W}^{3}_\mu)$ are again given by equations (\ref{covariant-1})$\sim$(\ref{W-bar}). The coupling constants $(g_3,g_0)$ are the same as eq.(\ref{couplings}). The rotation of gauge fields in this case, however, is defined by 
\begin{equation}
A =\bar{W}^3\cos\theta-W^0\sin\theta,~~~
Z =\bar{W}^3\sin\theta+W^0\cos\theta, \label{rotation-3}
\end{equation}
in such a way that $\bar{W}^3_\mu$ tends to $A_\mu$ in the limit $\theta\rightarrow 0$. The $\bar{W}^3,W^0$ terms in eq.(\ref{covariant-2}), then, can be written as
\begin{equation}
 g_3 \bar{W}_\mu^{3}{\cal J}_3+g_0W_\mu^0 S=eQA_\mu+\frac{e}{\sin\theta\cos\theta}\left\{ \frac{\alpha}{2}J_0+Q\sin^2\theta \right\}Z_\mu, \label{A-Z-2}
\end{equation}
providing $e=g_3\cos\theta$ and
\begin{equation}
\tan\theta=\frac{g_3}{g_0}\frac{\alpha}{Y}=\frac{\sqrt{\frac{N_0}{N_3}}\frac{\alpha}{2}}{\sqrt{1+\frac{N_0}{N_3}\left(\frac{\alpha}{2}\right)^2}}.
\end{equation}
Therefore, the mixing between $\bar{W}^3_\mu$ and $W^0_\mu$ is a direct result of a non-trivial deformation $q\neq 1$ in this case.

From these equations, one can evaluate the masses of vector bosons $M_\pm, M_Z $ caused by the symmetry breaking $\langle \Phi\rangle_0=(0,v,0)^T$. Using $\alpha=2-(q+q^{-1}), N_0=6,N_3=4, N_{\pm}=2(q+q^{-1})$, and $\{J_{+},J_{-}\}\langle\Phi\rangle_0=2(q+q^{-1})\langle\Phi\rangle_0$ for $J=1$, it is not difficult to verify that
\begin{equation}
 M_Z=\frac{g_3 v}{\sin\theta}\frac{\alpha}{2}=\frac{gv}{\sqrt{6}}\sqrt{1+\frac{3}{2}\left(1-\frac{q+q^{-1}}{2}\right)^2}
 ~~~{\rm and}~~~
 M_W=\frac{gv}{\sqrt{2N_\pm}}\sqrt{q+q^{-1}}=\frac{gv}{2}.
\end{equation} 
This leads to an undesirable result $M_W/M_Z\simeq 1.2$ for $q \rightarrow 1~(\theta \rightarrow 0)$, which may not be surprising, since non-trivial mixing $\theta\neq 0$ arises only for $q\neq 1$ in this case. In order to obtain the inequality $M_W/M_Z<1$, we have to require fairly large deformation $q>2.8$ or $0<q<0.36$. 

Finally, we comment on the Yukawa interaction term between the leptonic fields and the Higgs fields and the bilinear term of leptonic fields that are given  by
\begin{equation}
 {\cal L}_{\phi,\psi}=G_\Phi\{(\bar{\Psi}_R\vec{\cal J}\Psi_L)\cdot\vec{\Phi}+h.c. \}+M\bar{\Psi}\Psi, \label{Yukawa}
\end{equation}
where $\vec{\cal J}\cdot\vec{f}=\frac{1}{\sqrt{2}}({\cal J}_{+}f^{+}+{\cal J}_{-}f^{-})+{\cal J}_{3}f^3$. The interaction Lagrangian ${\cal L}_{\phi,\psi}$ are invariant obviously under $\star-$transformations caused by $({\cal J}_{\pm},{\cal J}_3)$; that is, $\delta_\star{\cal L}_{\phi,\psi}=0$. In particular, the ${\cal L}_{\phi,\psi}$ is invariant under a physical $U(1)$ charge transformation by $Q$. It should be noticed, however, that the both terms in ${\cal L}_{\phi,\psi}$ are not invariant simultaneously under the $U(1)$ transformation by $\hat{S}$, since a non-zero $U(1)$ hypercharge $Y_{H}$ is assigned for the Higgs field. Furthermore, one can verify that the ${\cal L}_{\phi,\psi}$ generates the mass terms $m_{+}\bar{\mu}{\mu}+m_{-}\bar{e}e$ with $m_{\pm}=M \pm G_{\Phi} v\sqrt{q+q^{-1}}$ after the symmetry breaking by $\langle \Phi\rangle_0=(0,v,0)^T$.

\section{Summary and Discussion}

In this paper, we have discussed a possible way to construct the electroweak gauge theory based on the $SU_q(2)$ symmetry. In the usual q-gauge theories, the gauge fields become non-commutative or non-local one's. The basic idea to ged rid of these problems is to introduce the $\star$-product such as $J_i\star J_j \equiv R_{ij}^{kl}J_k J_l$, where $R_{ij}^{kl}$ is a function of $J_3$. In other words, the $\star$-product is a kind of the redeformation of q-deformed algebra so as to recover the algebra before deformation under this product. Since, then, the $\star$-commutator losses the traceless property because of $Tr(J_i\star J_j)\neq Tr(J_j\star J_i)$, the $SU_q(2)$ generators form a closed algebra of the $SU(2)$ symmetry incorporated with a $U(1)$ generator in a non-trivial manner; in this sense, the gauge symmetry is written as $[SU_q(2)\times U(1)]_\star$. 

In the resultant $[SU_q(2)\times U(1)]_\star$ symmetric gauge theoreis, however, the $SU(2)$ symmetry is already broken in addition to the $SU_q(2)$ symmetry. Indeed, the trace of generators corresponding to $SU(2)$ symmetry $Tr({\cal J}_i^2), (i=\pm,3,0)$ are different each other because of their $q$-dependence.

According to this approach to $[SU_q(2)\times U(1)]_\star$ gauge theory, the gauge fields are sufficient to be ordinary commutative four-component one's. Further, since the formulation can be started with one gauge coupling constant, the Weinberg angle $\theta_W$ is determined uniquely for a given $q$; and, we tried two simple cases of matter fields belonging respectively to $J=\frac{1}{2}$ and $J=1$ representations.

The $J=\frac{1}{2}$ representation, the first case, is an exceptional case; then, the generators of $[SU_q(2)\times U(1)]_\star$ are reduced to those of $SU(2)\times U(1)$, to which the $q$-dependence is disappear. If we realize this model as a limiting case $J\rightarrow \frac{1}{2}$, the $\theta_W$ is determined uniquely; the value $\theta_W$ comes to be independent of $q$, though the result is not suitable for a phenomenology. On the other hand in the second case, a model of Higgs fields gives rise to the mixing angle $\theta$ and the ratio $M_W/M_Z$ that are determined depending on $q$. In order to obtain a physical ratio $M_W/M_Z<1$, we have to require a larger deformation such as $q>2.8$ or $0<q<0.36$. The $J=1$ model may be a special case of the triplet lepton fields tried by many authors\cite{triplet}; if we introduce an another neutral Higgs field $\bar{\phi}^0$ associated with the generator ${\cal J}_0$, the present formulation will close to those models.

The q-deformed gauge theory in this attempt is discussed within the framework of a gauge coupling between the gauge-Higgs contents $(W,\phi)$ and a matter field belonging to a irreducible representation of $SU_q(2)$ symmetry; if we consider a lager symmetry or a product symmetry, the situation for the parameters such as $\theta_W$ will be changed. Indeed, the addition of generators preserving $SU_q(2)$ algebra is realized in the Hopf structure in such a way that
\begin{align}
\Delta(J_3) &=J_3\otimes 1+1\otimes J_3 \\
\Delta(J_\pm) &=J_\pm \otimes q^{-J_3}+q^{J_3}\otimes J_\pm
\end{align}
The resultant representations are reducible; and so, the normalizations of generators are changed from original one's, although the $\eta$ opertor for $\Delta(J_3)$ and $\Delta(J_\pm)$ is again obtained by substituting $\Delta(J_3)$ for $J_3$ in eq.(\ref{eta-2}).

In this paper we confine our argument within the framework of classical field theories; in addition to those, the study of quantum correction in q-deformed gauge theories is also important future problem.

\subsection*{Acknowledgements}

The authors wish to express their thanks to the members of their laboratory for discussion and encouragement.

\appendix

\section{Representation of $SU_q(2)$ algebra}

We can construct the representation $SU_q(2)$ algebra in a similar way to the ordinary $SU(2)$ algebra. The difference lie only in the form of the second Casimir invariant $C_2$, which we can put without loss of generality as
\begin{equation}
 C_2=\frac{1}{2}\{J_{+},J_{-}\}+f(J_3), 
\end{equation}
where $f(J_3)$ is a function of $J_3$ determined by the requirement $[J_i,C_2]=0$. By definition, $[J_3,C_2]=0$ is satisfied obviously; and, further we have to put
\begin{equation}
 [J_{+},C_2]=\left[\frac{1}{2}\left([2J_3]+[2(J_3-1)]\right)+f(J_3-1)-f(J_3)\right]J_+ =0~,
\end{equation}
from which follows
\begin{equation}
 f(J_3)-f(J_3-1)=\frac{1}{2}\left([2J_3]+[2(J_3-1)]\right)~. \label{recurrence}
\end{equation}

Here, taking $[0]=0$ into account, the recurrence equation (\ref{recurrence}) can be solved easily as $f(J_3)=\frac{1}{2}[2J_3]+[J_3][J_3-1]+f(0)$. Requiring further $f(J_3) \rightarrow J_3^2,~(q \rightarrow 1)$, we obtain $f(0)=0$; then, the second Casimir invariant $C_2$, the $\bm{J}^2$ in Eq.(\ref{Casimir}), is decided as
\begin{align}
 C_2=\bm{J}^2 &=\frac{1}{2}\{J_{+},J_{-}\}+\frac{1}{2}[2J_3]+[J_3][J_3-1]
  \nonumber \\
              &=J_{+}J_{-}+[J_3][J_3+1]=J_{-}J_{+}+[J_3][J_3-1], \label{C2}
\end{align}
where we have used the relation $[2n]+[n][n-1]=[n][n+1]$. Therefore, the representation bases of $SU_q(2)$ algebra are characterized by the eigenvalue equations 
\begin{align}
 \bm{J}^2|\lambda,M \rangle &= \lambda|\lambda,M \rangle~, \label{eigen-3} \\
 J_3|\lambda,M \rangle &= M|\lambda,M \rangle~. \label{eigen-4}
\end{align}
A little calculation leads to the positivity of $\bm{J}^2$; that is, $\lambda \geq 0$. Further, one can verify that $J_{\pm}|\lambda,M \rangle \propto |\lambda,M \pm 1 \rangle$. Thus there are $J={\rm max}(J_3)$ and $\bar{J}={\rm min}(J_3)$ satisfying
\begin{align}
 \langle \lambda,J|J_{-}J_{+}|\lambda,J \rangle &= \lambda-[J][J+1] = 0, \\
 \langle \lambda,\bar{J}|J_{+}J_{-}|\lambda,\bar{J} \rangle &= \lambda-[\bar{J}][\bar{J}-1] = 0,
\end{align}
from which we have $\bar{J}=-J$ because of $[M][M+1]=[(-M)][(-M)-1]$. Thus we can write $\lambda=[J][J+1]$ and $|\lambda,M \rangle =|J,M\rangle$; then,  Eqs.(\ref{eigen-3}) and (\ref{eigen-4}) are nothing but Eqs.(\ref{eigen-1}) and (\ref{eigen-2}). With this eigenvalue of $\bm{J}^2$ and the normalization $\langle J,-J|J,-J\rangle=1$, the Eq.(\ref{C2}) gives rise to the normalization of eigen state $|J,M \rangle\propto (J_{+})^{J+M}|J,-J\rangle$ by
\begin{align}
 \| (J_{+})^{J+M}|J,-J\rangle \|^2 &= \langle J,-J|(J_{-})^{J+M}(J_{+})^{J+M}|J,-J\rangle \nonumber \\
 &= ([J][J+1]-[M][M-1])\| (J_{+})^{J+M-1}|J,-J\rangle \|^2 \nonumber \\
 &\vdots \nonumber \\
 &= ([J][J+1]-[M][M-1])!~.
\end{align}
Here, the $(\cdots)!$ implies the product with respect to eigenvalues of $J_3$ from $-J$ to $M$. The result backs up the form of normalized eigenstate $|J,M\rangle$ in E.(\ref{normalization}).

\subsection*{Acknowledgements}

The text of acknowledgements should be typed at the end of the paper, before references.

\end{document}